\documentstyle[12pt]{article}
\input psfig

\def\|{\'\i }
\def \nl{\noindent}

\begin{document}
\begin{center}{ \large {\bf TWO-LOOP CHIRAL 
PERTURBATION THEORY\\  AND THE PION-PION PHASE SHIFTS}}

\centerline{Submited  to Phys. Lett. B}  
\vspace*{.7 cm}

{J. S\'a Borges, J. Soares Barbosa and M. D. Tonasse \\
{ Instituto de F\|sica\\ Universidade do Estado do Rio de Janeiro}\\
Rua S\~ao Francisco Xavier 524, Maracan\~a \\Rio de Janeiro, Brasil \\
{J. Haidenbauer\\  Institut f\"ur Kernphysik, Forschungszentrum J\"ulich GmbH,
\\ D-52425 J\"ulich, Germany}}   
\end{center}
\pagestyle{empty}
\begin{abstract}

We want to test the predictive power of Chiral Perturbation Theory (ChPT).
In this work, we use the ChPT pion-pion scattering amplitude,  
including two loop contributions, and we obtain S- and P-wave low-energy 
phase shifts. We show that, by varying {\it just one} free parameter, 
the resulting S- and P-wave phase shifts are in reasonable agreement with 
the experimental data. 
\vskip .5 cm
\centerline{Pacs:  11.30Rd, 12.39.Fe, 13.75.Lb}

\centerline{ Key-words: Pion-pion Scattering; Chiral Symmetry; 
 Chiral Perturbation Theory.}
\end{abstract}

\section{Introduction}

The method of Chiral Perturbation Theory (ChPT)~\cite{Leut1} aims to
explore the low energy structure of Quantum Chromodynamics (QCD).
Considering a simultaneous expansion of the generating functional in
terms of the quark masses and momenta, this method gives the low
energy expansion of QCD Green's functions and in particular
meson-meson scattering amplitudes.

It is known that tree level ChPT calculations are equivalent to the
current algebra low-energy theorems and in particular that it reproduces
the pion-pion Weinberg amplitude~\cite{Wei1}. On the other hand, loop
diagrams in ChPT give quite large corrections to leading current
algebra results even at threshold.

The other method to describe low energy meson-meson scattering was
invented in the early sixties - the hard-meson current algebra technique.
Even ignoring the underlying theory, the chiral current algebra
implies a set of Ward identities and this method consists in solving
the system of Ward identities under suitable assumptions~\cite{Wei2}.

In this context we have introduced unitarity corrections to current
algebra soft-meson result and we call this approach {\em
unitarization program of current algebra} (UPCA). The application of
UPCA allows one to go beyond threshold of meson-meson scattering and to
access even the resonance region.

As ChPT and UPCA follow from chiral symmetric Ward identities we
were interested in comparing their results. We have compared these
two methods for the one-loop ChPT pion-pion and kaon-pion 
amplitudes~\cite{Bor1} and for the two-loop ChPT pion-pion 
amplitude~\cite{Bor2}. From this comparison we have concluded that UPCA
quasi-unitarized amplitudes, published long time ago \cite {Bor3},
have the same analytical structure as the corresponding ChPT ones:
They have the same dependence on $s$, $t$ and $u$, they are crossing
symmetric and they respect approximate unitarity for partial waves.
The expressions for the amplitude differ only in their energy polynomial parts
because the free parameters have different origins:
In the UPCA approach they are related to subtraction constants, inherent
to the dispersive technique, and in ChPT they are coupling
constants of the Lagrangian.

From the very beginning of the application of ChPT to describe
meson-meson scattering it was clear that one could not relate all
its free parameters to the QCD scale and to the quark masses. 
Like in UPCA, some have to be obtained phenomenologically. Certain 
constraints like the experimental value of the D-wave pion-pion 
scattering length and the electromagnetic charge radius of the pion 
were used to fix these parameters, leaving however still an
uncertainty range of 40\% to 60\%. We have
been interested in the possibility of fixing these free parameters by
fitting low-energy meson-meson phase shifts. Analyzing phase shifts based on
ChPT amplitudes calculated at next-to-leading order, we have
fixed \ $\bar \ell_1$\ and\ $\bar\ell_2$ by fitting pion-pion
experimental data and we have fixed \ $L^r_1,L^r_2$\ and\
$ L^r_3$ by fitting kaon-pion experimental data~\cite{Bor4}.

Recently the complete \ $O(p^6)$\ ChPT pion-pion amplitude was 
determined~\cite{Eck1}. The final expression contains six linear
combinations of ten free parameters. Therefore the question
concerning the predictive power of such a result arises. One 
possibility is to consider some estimates for relevant low-energy 
constants of $O(p^6)$ from another effective theory\cite{Eck2}, 
to keep \ $O(p^4)$\ constants in the same range of values
previously used and to assume that ChPT is restricted to very 
low energies. The results in Ref.~\cite{Eck1} for two different 
scales are consistent with 
data but the predictive power of the theory remains questionable.

Here we present a different proposal. We would like to test the
predictive power of ChPT by using the smallest possible number of parameters. 
Thereby, we deal only with {\it analytical expressions} for the
partial waves and we try to describe pion-pion scattering by varying
{\it just one} parameter. As a result, we
obtain S- and P-wave phase shifts that are in reasonable agreement with
experimental data. We interpret the ability of ChPT to describe
these phase shifts, once given the \ $\rho$\ mass, as an 
indication that its predictive power does not depend on the large number 
of free parameters. In the next section we present the 
ChPT pion-pion amplitude, the phase-shift definition, 
numerical results and an analysis of the near-threshold behavior. 
In the conclusions we include a short discussion concerning the 
D-wave amplitude.

\section{ ChPT amplitudes and numerical results}

The  amplitude for elastic pion scattering
obtained from ChPT Lagrangian \cite{Eck1}, including two loop contributions 
with \, $m_\pi = 1 $\ and $F_\pi = 93.2/140$\,  is:

\begin{eqnarray}
 A^{(1)}(s,t,u)&=&(s-1)/F_\pi^2 +  \cr &&
\left[b_1+b_2 s  + b_3 s^2 
+ b_4 ( t - u ) ^2\right]/F_\pi^4 + \cr &&
\left[F^{(1)}(s) + G^{(1)}(s,t) + G^{(1)}(s,u)\right]/F_\pi^4
+ \cr && \left[ b_5 s^3 + b_6 s (t-u)^2\right]
/F_\pi^6
\cr &&\left[F^{(2)}(s) + G^{(2)}(s,t) + G^{(2)}(s,u)\right]/F_\pi^6
\label{1}\end{eqnarray}
\nl  where: 
\begin{eqnarray*}
  F^{(1)}(s)& = & \frac{1}{2} \bar J(s)\left( s^2 - 1\right),\\ 
F^{(2)}(s) & = & \bar J(s)\left \{\frac{1}{16\pi^2} \left ( \frac{503}{108} 
s^3 - \frac{929}{54} s^2 +\frac{887}{27} s -\frac{140}{9}\right) + 
b_1 ( 4 s - 3 ) + b_2 (s^2 + 4 s - 4)\right. \cr
&& \left. +   \frac{1}{3} b_3 ( 8 s^3 - 21 s^2 + 48 s - 32) + \frac{1}{3} 
b_4 ( 16 s^3 - 71 s^2 + 112 s - 48 ) \right \}\cr
&&  + \frac{1}{18} K_1(s)  \left [ 20 s^3 - 19 s^2 + 210 s -135
-\frac{9}{16}\, \pi^2 ( s - 4 ) \right ] \cr
&&   + \frac{1}{32} K_2(s) \left ( s \pi^2 -24 \right ) +\frac{1}{9}
K_3(s) \left ( 3 s^2 - 17 s + 9 \right ),\cr 
 G^{(1)}(s,t)& = & \frac{1}{6} \bar J(t) \left(14 - 4 s  - 10 t  
+ s t + 2 t^2\right),\cr
G^{(2)}(s,t) & = & \bar J(t) \left \{ \frac{1}{16 \pi^2}
\left [ \frac{412}{27} - \frac{ s}{54} ( t^2 + 5 t + 159 ) - t \left(
\frac{267}{216} t^2 - \frac{727}{108} t + \frac{1571}{108} \right) \right ]
\right.  \cr
&&    + b_1 ( 2 - t ) +\frac{1}{3} b_2 ( t - 4 ) ( t^2 + s - 5 ) -
\frac{1}{6} b_3 ( t - 4 )^2 ( 3 t + 2 s - 8 ) \cr
&& \left. + \frac{1}{6} b_4 ( 2 s ( 3 t - 4 ) ( t - 4 ) - 32 t + 40 t^2 -
11 t^3) \right \}  \cr
&&   + \frac{1}{36} K_1(t) \left [ 174 + 8 s - 10 t^3 + 72 t^2 - 185
t - \frac{1}{16}\, \pi^2 ( t - 4 ) ( 3 s - 8 ) \right ]  \cr
&&    + \frac{1}{9} K_2(t) \left [ 1 + 4 s + \frac{1}{64} \, \pi^2 t (
3 s - 8 ) \right ]  \cr
&&    + \frac{1}{9} K_3(t) \left ( 1 + 3 s t - s + 3 t^2 - 9 t \right
) + \frac{5}{3} K_4(t) \left (4 - 2s - t \right ).
\end{eqnarray*}

\noindent
In this expression 
\begin{eqnarray*}
16\pi^2 \bar J (s)&=\sigma(s)
h(s) + 2,\quad 
(16\pi^2)^2\, K_1(s)   = h^2(s),\quad  (16\pi^2)^2\, K_2 (s) =  \sigma^2(s)\, 
 h^2(s) - 4, \cr \\ 
 (16\pi^2)^2\, K_3& =\frac {1}{s\, \sigma}\, 
h^3(s) + \frac{\pi^2}{s\, \sigma} h(s) - \frac{\pi^2}{2}\quad  {\hbox{and}}\quad
  K_4 =  
\frac{1}{s\, \sigma^2}\left ( \frac{1}{2}\, K_1 + \frac{1}{3} K_3 +
\frac{1}{16\pi^2} \, \bar J + s\,  \frac{ \pi^2 - 6 }{192\pi^2}\right ),
\end{eqnarray*}
\nl where
$$h(x) = \ln\  \frac{\sigma(x) - 1  }{\sigma(x) + 1}\quad{\hbox{with}}
\quad \sigma(x) = \sqrt {1 - \frac{4}{x}}.\eqno{(2)} $$
The total amplitudes $T_I$ for the isospin \ $I=0, \, 1\ {\hbox{and}}\  2$ 
channels are :
$$T_1(s,t)= A(t,s) - A(u,t) , \ T_2(s,t) =A(t,s) + A(u,t)\quad 
{\hbox{and}}\quad  T_0 (s,t) = 3 A(s,t) + T_2(s,t).$$
We work with S- , P- and D-waves obtained from the total amplitudes by:
$$t_{I\, \ell} (s) = \frac{1}{64\pi} \int_{-1}^1 T_I(s,t,u) P_\ell(x) dx,$$
\nl where\quad $2 t = ( s - 4) ( x - 1 )$, \quad $ 2 u = - ( s - 4) ( x + 1 )$ 
\ and \   $P_\ell$\  are Legendre polynomials.

We take this opportunity to introduce some of the usual phase shift definitions 
in order to establish the difference between the one employed in the present 
paper and that used in other approaches.

It is well known that, in the elastic region \ $4 < s < 16 $,\ unitary 
implies that the partial wave amplitudes \ $t_{I\, \ell}$\ fulfil
the relation:
$$  I_m\  t_{I\, \ell}(s) = \frac{\sigma}{32 \pi}\ \vert\ t_{I\, \ell}(s)\
\vert\ ^ 2. $$ 
This relation allows one to define 'exact' phase shifts 
$\delta_{I\, \ell} $ so that 
$$  t_{I\, \ell}  (s) = \frac{32 \pi}{\sigma}\ 
exp\ \left (i\ \delta_{I\, \ell} (s)\right)\  \sin \ \delta_{I\, \ell} (s),
 $$

\nl However, ChPT partial waves do not satisfy exact elastic unitarity 
and, for this reason, some authors \cite{Eck1} introduced another
definition, namely
$$ \delta_\ell^I(s) = \frac{\sigma}{32 \pi} R_e\  t_\ell^I(s),$$

\noindent  which is a good approximaton to the 'exact' definition
for small values of $\delta_{I\, \ell}$. 

In order to exploit the right-hand-side discontinuity of the amplitude 
generated by the loop calculation, leading to an imaginary part, we have 
adopted another definition for $\delta_{I\, \ell}$ in applications
of the UPCA~\cite{Bor3} which we also use in the present analysis, namely:
$$ \delta_{I\, \ell}(s)\ = \ \tan ^{-1}\   \frac {I_m\
t_{I\, \ell}(s)}{R_e\ t_{I\, \ell}(s)}.\eqno {(3)}$$

In this paper we fix almost all parameters  to be zero and we vary
{\it just $b_4$}. We show the S- and 
P-phase shifts in Fig. 1 and Fig. 2 respectively for the 
{\it one-loop} amplitude with $b_4 =-.005$
and in Fig. 3 and Fig. 4 respectively for the 
{\it two-loop} amplitude with $b_4 = .025$.

To analyze our results it is convenient to divide the total amplitude (
Eq.(\ref {1})) in three parts: A first part which 
contains the function $h(s)$\ up to power three;
a second part which contains $h(t)$ (or $h(u)$)\ up to the third 
power and a rest which then contains neither $h(s)$ nor $h(t)$. 
This last part includes the lowest order term in the chiral expansion 
as obtained by Weinberg~\cite{Wei1}.

$h(s)$ is analytic in the complex s-plane cut along the positive 
real axis in the physical region ($s\geq 4 $), therefore s-channel 
contributions corresponding to powers of $h(s)$ will give rise to a real
($t^{right}$) and an imaginary 
part ($t^{imag}$) for the partial wave amplitude.
Crossed channel contributions, corresponding to the integration 
of $h(t)$ and its powers, give rise to functions that are discontinuous on 
the left hand side ($s\leq 0$) and we will call this part $t^{left}$. 
The part of the amplitudes that contains neither  
$h(s)$ nor $h(t)$\ gives rise to a polynomial contribution. This part
includes the Weinberg amplitude, which we denote by $t^{ca}$, 
and a remainder which we call $t^{free}$. Needless to say that the
parameters $b_i$ appear in several parts of the amplitudes.

It is very instructive to know how each part of the amplitudes behaves 
near threshold when compared with the Weinberg amplitude. We show in Fig. 5 
(for P-wave) and Fig. 6 (for S-wave) the ratio of each contribution to 
the soft-pion result \ $t^{ca}$\ (straight lines), for energies from 
280 to 500 MeV. 
Curve (a) is the ratio $t^{free}$/$t^{ca}$, curve (b)
is the ratio $t^{right}$/$t^{ca}$ and curve (c) 
is the ratio $t^{left}$/$t^{ca}$. These figures help us to 
see that the leading contribution is $t^{ca}$, but the corrections 
start to be important for energies bigger than 500 MeV.

At this point we would like to discuss another consequence of \ $O(p^6)$\ 
ChPT calculation. The UPCA second order corrected partial waves exhibit 
a discontinuity as follows:

$$ {\hbox{Im}}\ t^{(2)}_{\ell I}(s) = \frac{1}{16\pi} \sigma (s)\ 
t^{ca}_{\ell I}(s) {\hbox{Re}}\ t^{(1)}_{\ell I}(s)\quad
{\hbox{for}}\quad s\geq 4 .$$

In this expression the superscripts indicate the order of the approximation.
Since the Weinberg amplitude is linear in \ $s$, there is no soft pion 
contribution ($t^{ca}$) for the D-wave and we conclude that 
the resulting D-wave amplitude is real for\  $s\geq 4$. 
Now, since we have proven that the second order unitarity correction to
current algebra by UPCA is equivalent to $O(p^6)$ ChPT, this means that a 
two-loop calculation from ChPT does not provide any D-wave phase shifts
using our definition. 
In principle, ChPT \ $O(p^8)$\ calculation allows a complex amplitude 
for the D-wave but we guess that the task of constructing this amplitude 
can be easier done by the UPCA method~\cite{Bor1}.

\begin{figure}[htbp]
\psfig{figure=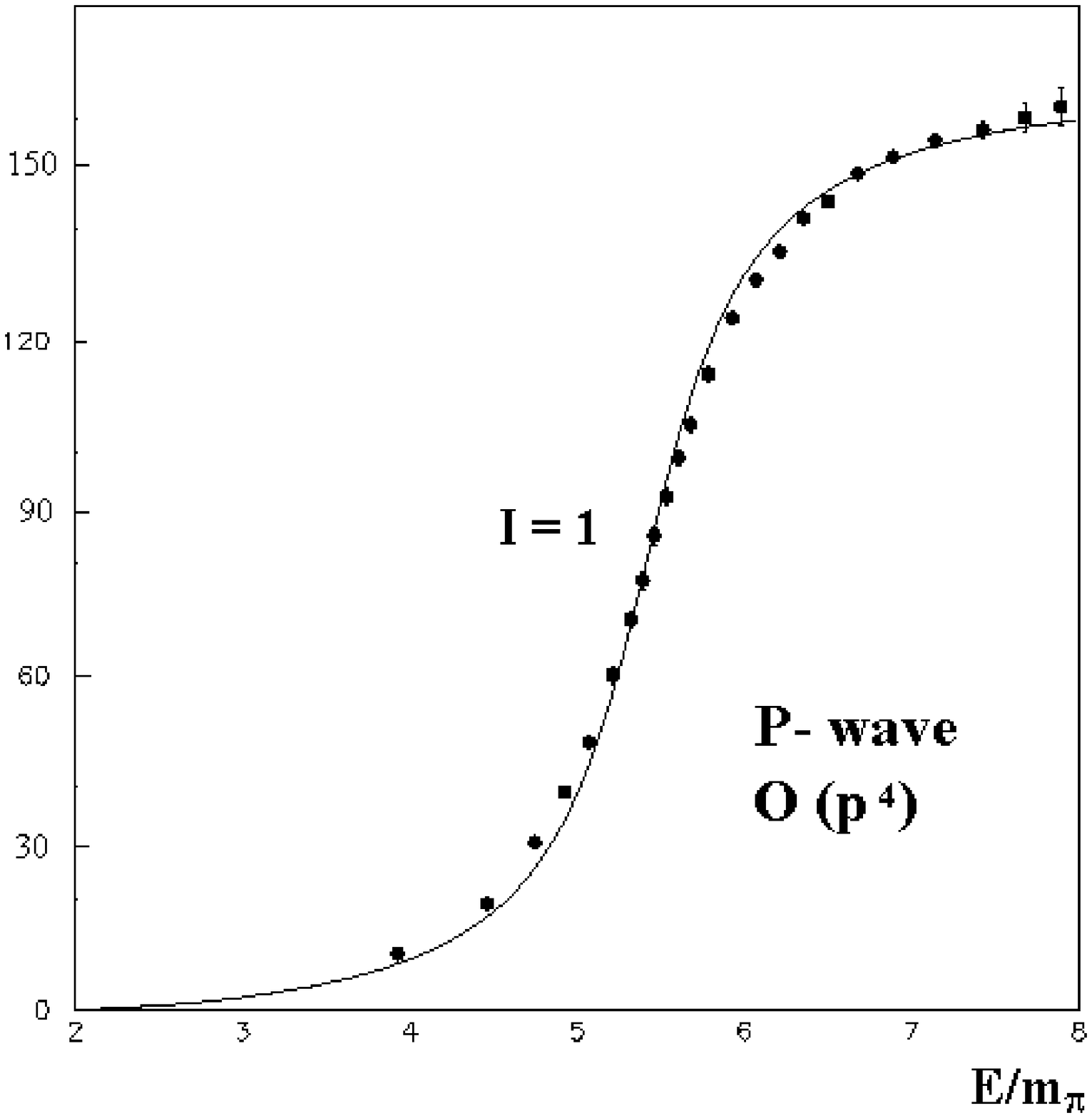,height=6cm,width=8cm}
\vspace*{-6cm}
\hspace*{4cm}
\centerline{\psfig{figure=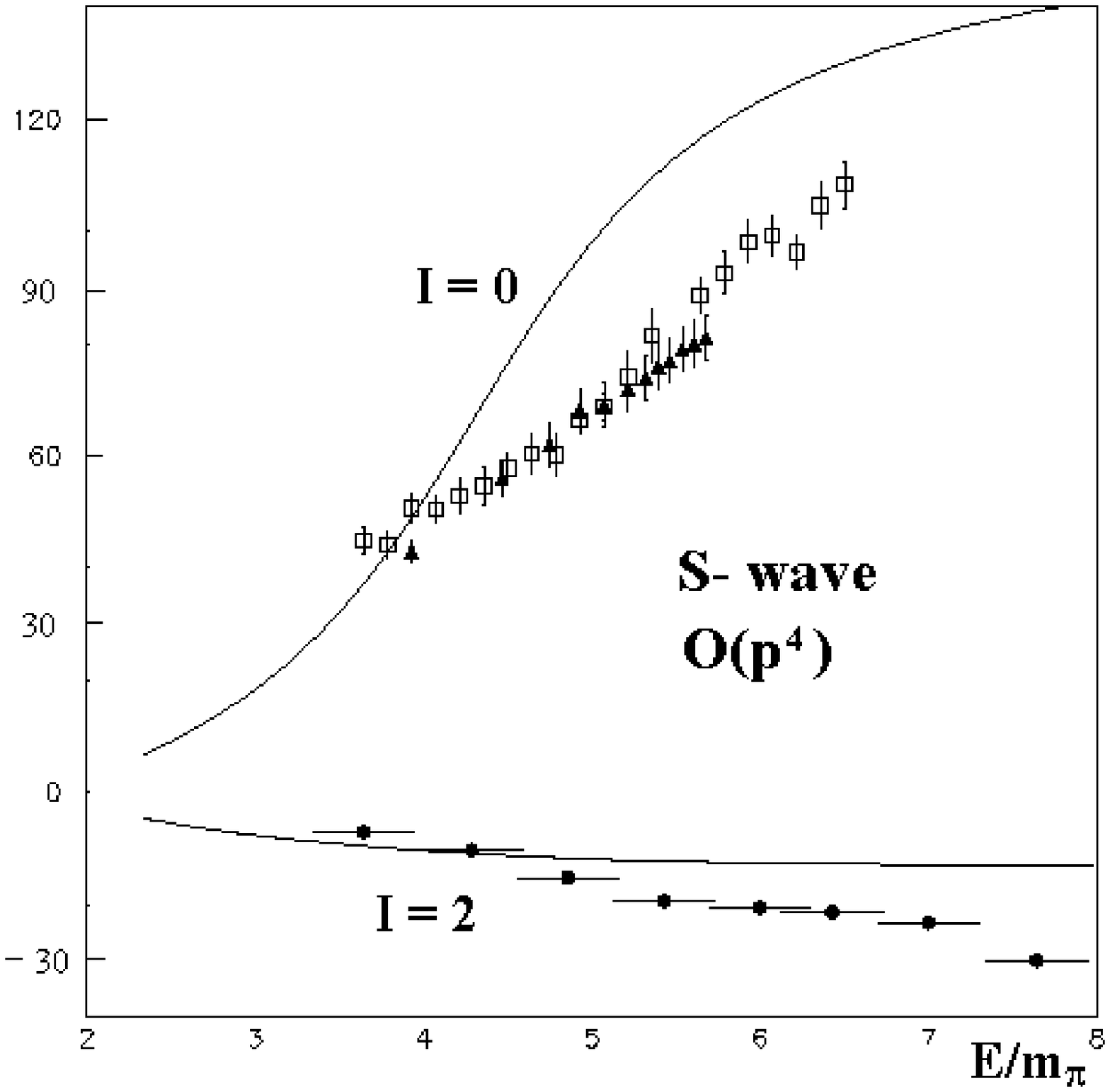,height=6cm,width=8cm}}
\parbox{7.5cm}{FIG. 1. $O(p^4)$\ ChPT P-wave phase shifts 
with\ 
$b_4 = -.005$ and experimental \cite{Pro} data.}
\hspace*{1.2cm}
\parbox{7.5cm}{FIG. 2. $O(p^4)$\ ChPT S-wave phase shifts 
with\ $b_4 = -.005$ and experimental\cite{Pro,Est}  data.} 
\end{figure}

\begin{figure}[htbp]
\psfig{figure=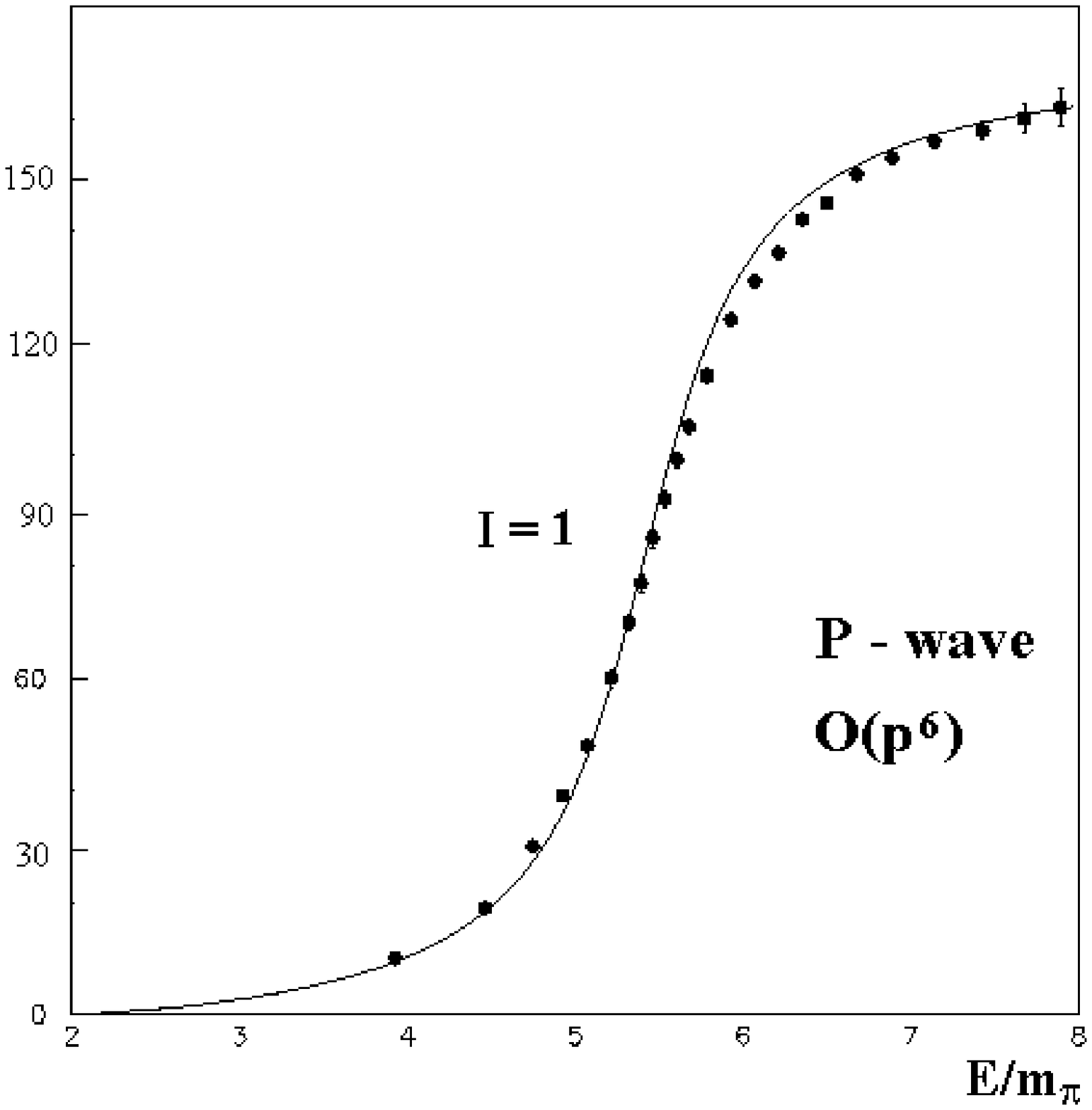,height=6cm,width=8cm}
\vspace*{-6cm}
\hspace*{4cm}
\centerline{\psfig{figure=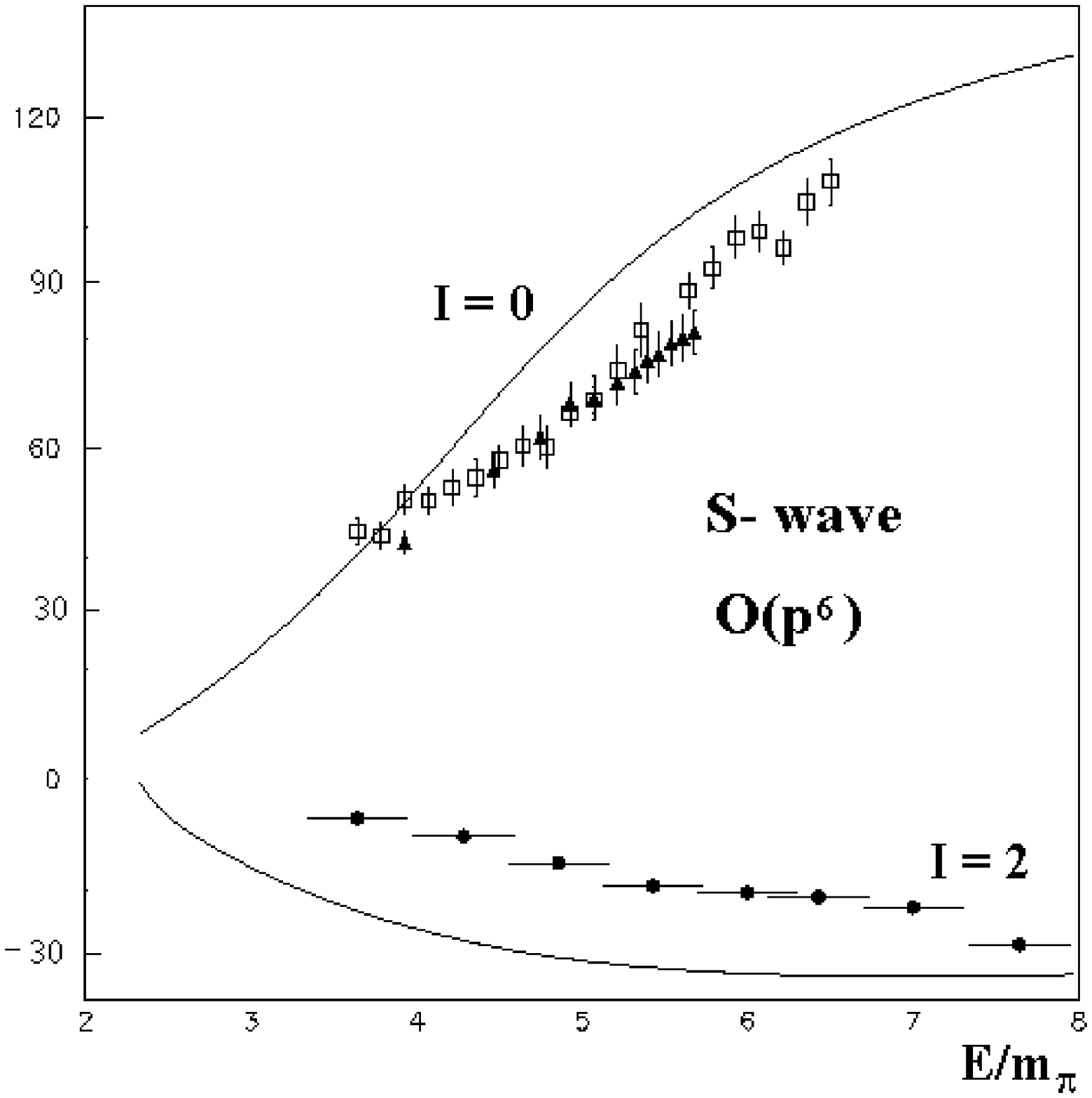,height=6cm,width=8cm}}

\parbox{7.5cm}{FIG. 3. $O(p^6)$\ ChPT P-wave phase shifts 
with\ 
$b_4 = .025$ and experimental \cite{Pro} data.}
\hspace*{.8cm}
\parbox{7.5cm}{FIG. 4. $O(p^6)$\ ChPT S-wave phase shifts 
with\ $b_4 = .025$ and experimental\cite{Pro,Est}  data.} 
\end{figure}

\begin{figure}[htbp]
\psfig{figure=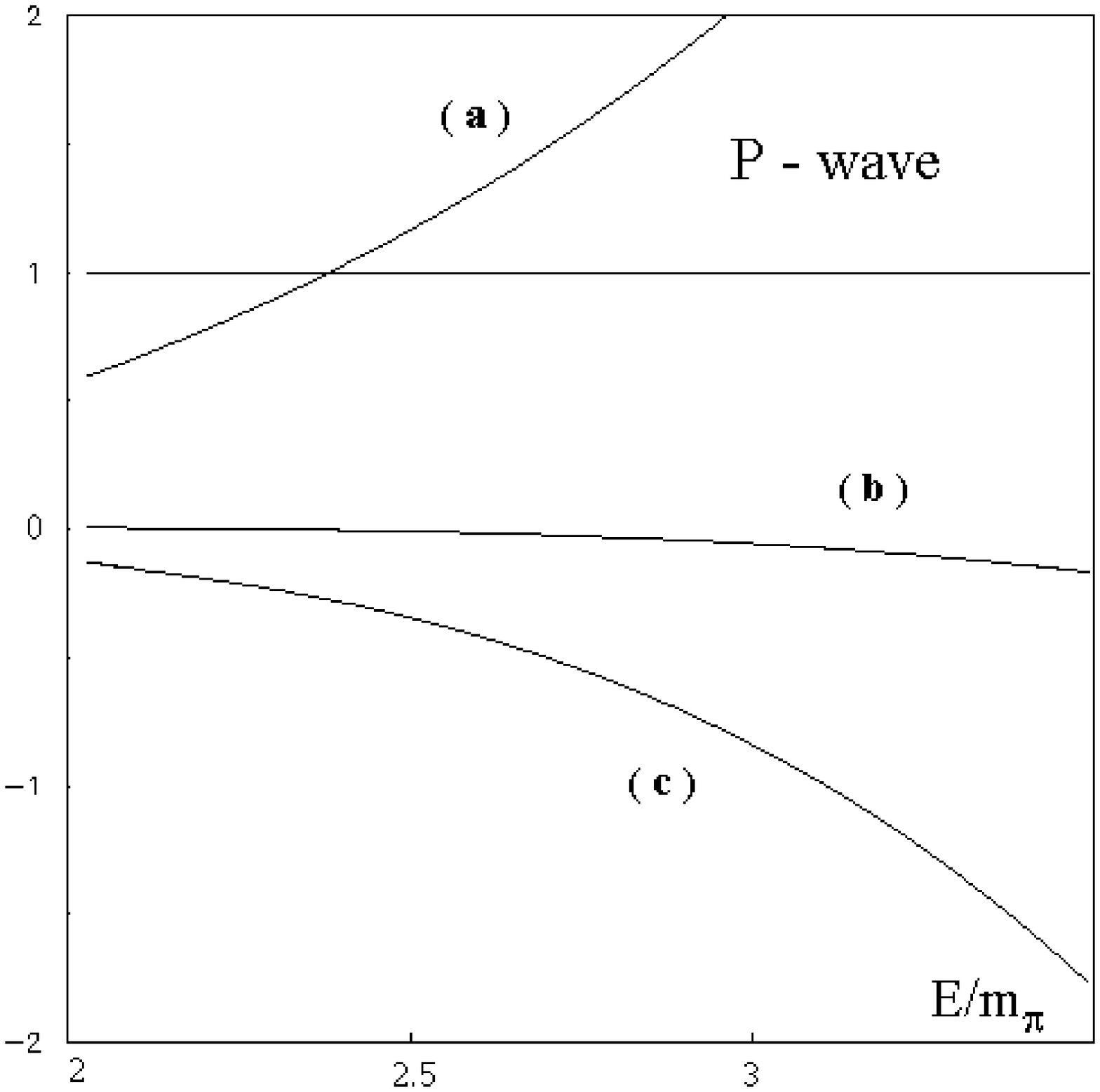,height=6cm,width=8cm}
\vspace*{-6cm}
\hspace*{4cm}
\centerline{\psfig{figure=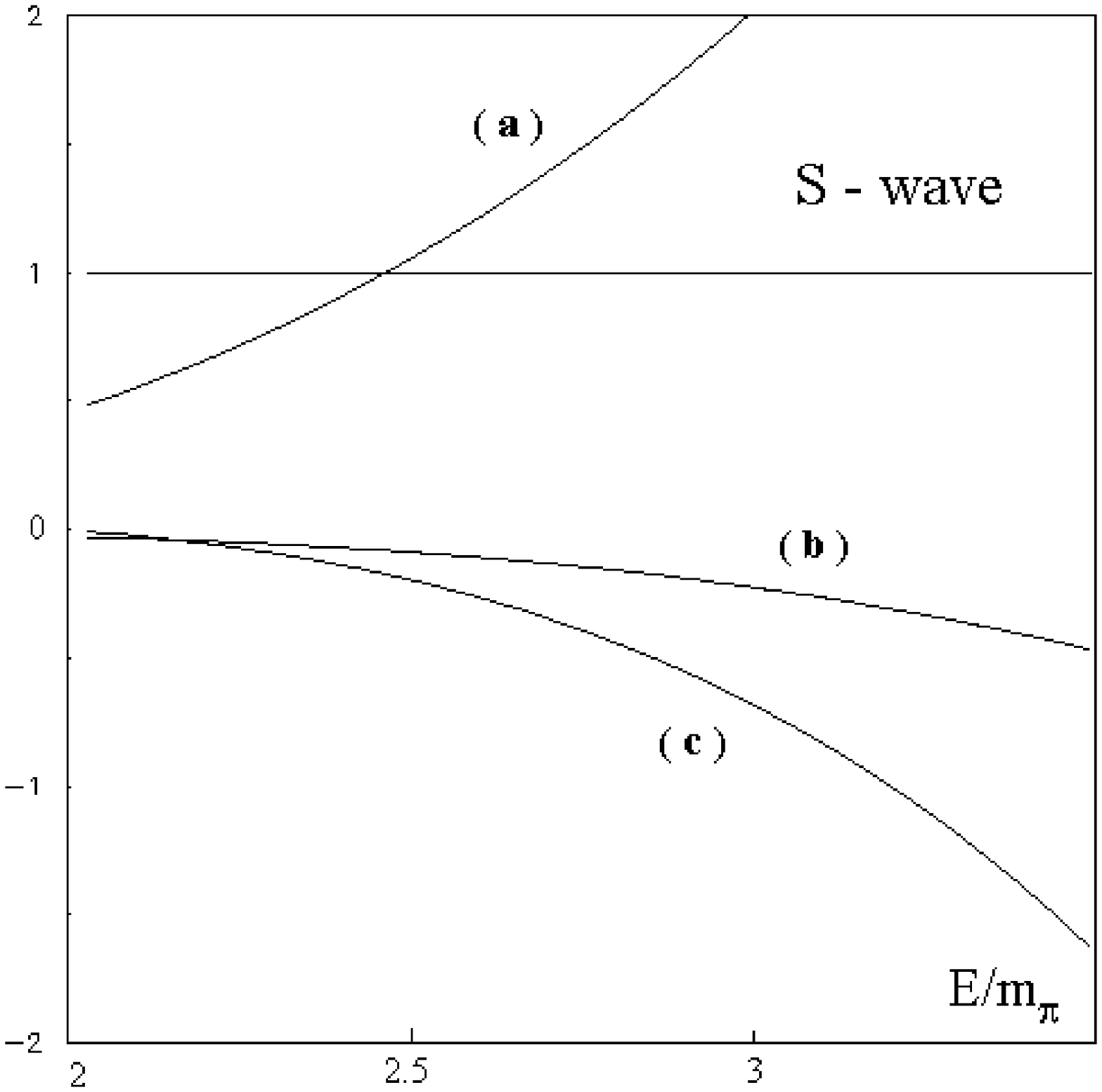,height=6cm,width=8cm}}
\parbox{7.5cm}{FIG. 5. Near the threshold behavior of P-wave 
amplitude components: Curve (a) is the ratio $t^{free}$/$t^{ca}$, 
curve (b) is the ratio $t^{right}$/$t^{ca}$ and curve (c) is the 
ratio $t^{left}$/$t^{ca}$.}
\hspace*{.8cm}
\parbox{7.5cm}{FIG. 6. Near the threshold behavior of S-wave 
amplitude components: Curve (a) is the ratio $t^{free}$/$t^{ca}$, 
curve (b) is the ratio $t^{right}$/$t^{ca}$ and curve (c) is the 
ratio $t^{left}$/$t^{ca}$. } 
\end{figure}

\section{Conclusions}
We have tested the predictive power of the ChPT amplitude for pion-pion
scattering.
We conclude that at $O(p^4)$\ and at $O(p^6)$,\  with {\it just one} 
free parameter, ChPT provides a qualitative description of this process. 
On the hand we are calling attention to the fact that, up to the two loop 
approximation, ChPT still does not provide an imaginary part for the D-wave
amplitude.
\newpage

\centerline{\bf ACKNOWLEGMENTS}

The work of M. D. Tonasse was supported by the Funda\c c\~ao de Amparo \`a 
Pesquisa no Estado do Rio de Janeiro (proc. E-26/150.338/97).
J. H. and J. S\'a Borges acknowledge financial support provided by a DLR (Germany)
- CNPq (Brazil) agreement, project no. BRA W0B 2F.

\end{document}